\documentclass[twocolumn,prc,aps,showpacs,amsmath,amssymb,nofootinbib]{revtex4}
\usepackage{graphicx}
 \newcommand\la{\langle}
 \newcommand\ra{\rangle}
 \newcommand\beq{\begin{equation}}
 
 \newcommand\eeq{\end{equation}}
 \newcommand\beqn{\begin{eqnarray}}
 \newcommand\eeqn{\end{eqnarray}}
 \newcommand\GeV{{\rm GeV}}
 
\def\fm{\,\mbox{fm}}
\def\GeV{\,\mbox{GeV}}

\def\lsim{\mathrel{\rlap{\lower4pt\hbox{\hskip1pt$\sim$}}
    \raise1pt\hbox{$<$}}}         
\def\gsim{\mathrel{\rlap{\lower4pt\hbox{\hskip1pt$\sim$}}
    \raise1pt\hbox{$>$}}}         

\def\la{\langle}
\def\ra{\rangle}

\begin{document}

\title{Survival of charmonia  in a hot environment}

\author{B.~Z.~Kopeliovich, I.~K.~Potashnikova, Iv\'an~Schmidt and M.~Siddikov}

\affiliation{\centerline{Departamento de F\'{\i}sica,
Universidad T\'ecnica Federico Santa Mar\'{\i}a; and}
Centro Cient\'ifico-Tecnol\'ogico de Valpara\'iso;
Avda. Espa\~na 1680, Valpara\'iso, Chile
}

\begin{abstract}
A colorless $\bar cc$ dipole emerging from a heavy ion collision and developing
the charmonium wave function can be broken-up by final state interactions (FSI) 
propagating through the hot medium created in the collision. We single out two 
mechanisms  of charmonium attenuation: (i) Debye color screening, called melting; 
and (ii) color-exchange interaction with the medium, called absorption. 
The former problem has been treated so far only for charmonia at rest embedded in the medium, while in practice their transverse momenta at the LHC are quite high, $\la p_T^2\ra=7-10\GeV^2$.
We demonstrate that a $\bar cc$ dipole may have a large survival probability even at infinitely high temperature. We develop a procedure of Lorentz boosting of the Schr\"odinger equation to a moving reference frame and perform the first realistic calculations of the charmonium survival probability employing the path-integral technique, incorporating both melting and absorption.  These effects are found to have comparable magnitudes. We also calculated the FSI suppression factor for the radial excitation $\psi(2S)$ and found it to be stronger than for $J/\psi$, except large $p_T$, where $\psi(2S)$
is relatively enhanced. The azimuthal asymmetry parameter $v_2$ is also calculated.
\end{abstract}



\maketitle

\section{Introduction: Melting vs absorption}

Charmonium production in heavy ion  collisions can serve as a hard QCD probe for the properties of the created dense medium, because the production time of a $\bar cc$ pair $t_{pr}\sim1/\sqrt{4m_c^2+p_T^2}$ is much shorter than the coherence time of most of the gluons radiated in a heavy ion collision.  
The properly normalized ratio $R_{AA}$ of the production rates in nuclear and in $pp$ collisions is  affected by the initial state (ISI) and final state (FSI) interaction stages, which experimentally are difficult to disentangle. The former corresponds to the early stage of the collision of the nuclei, which lasts a very short time interval,
$ t_{ISI}\sim 2m_NR_A/\sqrt{s}$,
where $R_A$ is the nuclear radius. The FSI stage lasts much longer, while the charmonium is propagating through the medium created in the collision, which has dimensions of the order of the nuclear radius, $t_{FSI}\sim R_A$.
A charmonium detected at the mid rapidity with transverse momentum $p_T$ can be treated in the c.m. frame as propagating with momentum $p_\psi=p_T$ through the medium. If charmonium is produced with nonzero rapidity, one can make a longitudinal Lorentz boost to the rest frame of the co-moving matter.

Charmonium attenuation in a hot matter is traditionally considered as a sensitive probe for  the medium temperature. If the Debye screening radius becomes smaller than the charmonium, the binding potential is expected to be screened, resulting in dissociation of the charmonium into an unbound $\bar cc$ pair. However, such a way to measure the medium temperature, proposed in \cite{satz}, has received no unambiguous confirmation so far, after two decades of experimenting. 

There are several reasons, which make difficult to single out the mechanism of charmonium melting \cite{psi-AA,puzzles}. First, charmonium  attenuation occurs already in the ISI stage, frequently labelled as cold nuclear matter effects. In spite of the pretty well developed dipole phenomenology of such effects \cite{rhic-lhc}, there is still no consensus regarding the interpretation of this stage. Even data on nuclear effects for charmonia produced in $pA$ collisions are not sufficient to disentangle the ISI stage, because no direct, model independent transition from $pA$ to $AA$, is possible \cite{nontrivial}.

Moreover, even if the ISI stage were absent, the interpretation of
charmonium attenuation in the final state interaction (FSI) stage would be far from being unambiguous.  
The available theoretical tools used to describe the dissolution of charmonia, look to us quite underdeveloped. The transverse momentum of $J/\psi$ detected at LHC ranges up to $100\GeV$ and its mean momentum squared varies with the collision energy between $7$ and $10\GeV^2$ \cite{psi-alice}.
However, the effects of melting has been calculated so far either assuming charmonium  being at rest relative to the medium \cite{satz,kks}, or embedded into a moving homogeneous plasma in a thermal bath of a constant temperature \cite{matsui,escobedo1,escobedo2}, which limits applicability of such a description to heavy ion collisions.

Besides, the charmonium is usually assumed to be either completely terminated \cite{kks}, once the bound level is eliminated by the screening, or the decay of the bound state is described by introducing an effective imaginary part of the potential, giving rise to a decay width \cite{laine}. 

Disappearance of a charmonium in the final state due to 
 Debye color screening  could indeed happen, if the charmonium were at rest in a static medium. However, most of charmonia observed experimentally are moving through the medium with relativistic velocities.
Such a charmonium has a finite probability to survive, even at infinitely high temperature, when the color screening completely eliminates the $\bar cc$ potential inside the medium, as is demonstrated in 
Sect.~\ref{maximal}.

Another important effect of FSI, widely missed in previous calculations, is a break-up of the colorless dipole caused by color-exchange interactions with the surrounding matter (see a short version of this paper in \cite{qm2014}).
Such a mechanism of attenuation has been known for hadrons propagating through a cold nuclear matter,  since the origin of the Glauber model, where it is usually called absorption, the term also used here.
A quantum-mechanical description of the propagation of color dipoles through a nuclear medium, the path-integral technique, was proposed in \cite{kz91,preasymptotics}. The imaginary part of the light-cone potential, responsible for absorption, was found to be proportional to
the dipole-proton cross section, which is well constrained by DIS data from HERA (see e.g. in\cite{joerg-hf}). Certainly, 
absorption becomes even stronger in a medium much denser than nuclei. However, the dipole cross section of interaction with the dense medium is unknown, cannot be fixed by other data.
The important observation, made in \cite{psi-AA}, is a relation between the imaginary part of the potential and the transport coefficient, which is related to the medium temperature (see below). The path-integral technique \cite{kz91} is presented in Sect.~\ref{path}. 

Notice that an attempt to implement absorption into the FSI stage,
relying on the path-integral technique of refs.~\cite{kz91,joerg-hf},  was made in \cite{blaschke}. However, the light-cone description, designed in \cite{kz91} for relativistic dipoles, was applied
in \cite{blaschke} to a charmonium at rest, where the imaginary part
of the potential, proportional to velocity, must vanish.

Imaginary part of the potential generated by the interaction with the surrounding medium
was also introduced in \cite{brambilla}, however this mechanism is significantly different from
ours, and the break-up cross section is much smaller. The transition of a color-singlet $\bar cc$ to a color-octet occurs due to direct absorption of a gluon from the medium \cite{peskin}.
The cross section of this process is steeply falling with the c.m. energy of $\psi$-$g$ collision,
as $(s_{\psi-g}-M_{\psi}^2)^{-7/2}$, while the cross section of $t$-channel gluon exchange,
considered here, is rising with energy. Besides, gluon absorption by an on-mass-shell quark is impossible, so the corresponding cross section steeply vanishes at small
binding energies, as $\epsilon^3$ \cite{peskin}. The latter is expected to be diminished in a hot medium due to Debye screening.

In order to make use of the potential and screening corrections, which are known in the charmonium rest frame, a procedure of a Lorentz boosting of the Schr\"odinger equation to a moving reference frame is developed in Sect.~\ref{boost}. It allows to find the Green function describing 
the propagation and evolution in the medium of a $\bar cc$ pair interacting with each other and with the medium. Numerical calculations of the survival probability of charmonia ($J/\psi,\ \psi(2S)$) and the azimuthal asymmetry parameter $v_2$ are performed in Sect.~\ref{results}.

The results are summarised in Sect.~\ref{summary}, and possible further development of 
the mechanisms of charmonium production in nuclear collisions are highlighted in Sect.~\ref{outlook}.

\section{Absorption in cold and hot media}\label{absorption}

It has been known since the origin of the Glauber model \cite{glauber} that the survival probability of a high-energy hadron propagating through
nuclear matter attenuates exponentially,
\beq
S^2(l_1,l_2)=\exp\left[-\sigma_{in}\int\limits_{l_1}^{l_2}dl\,\rho_A(l)\right],
\label{200}
\eeq
where the integration is performed along the straight hadron trajectory from the initial to final coordinates, $l_{1,2}$;
$\sigma_{in}$ is the hadron-nucleon inelastic cross section; $\rho_A(l)$ is the nuclear density.

The source of attenuation, the possibility of inelastic interactions of the hadron, is usually called absorption.
The inelastic interaction is a result of a color exchange in the cross channel. Thus, Eq.~(\ref{200}) describes the probability for the hadron to avoid color-exchange collisions, i.e. to remain colorless. This mechanism of attenuation is different from color screening in a medium, called melting or dissolving, considered in the next section.

It is convenient to relate the rate of inelastic collisions in the nucleus, $\sigma_{in}\rho_A(l)$, used in (\ref{200}), with the transport coefficient
$\hat q$, which is the rate of broadening of the transverse momentum $k_T$ of a parton propagating through the medium \cite{bdmps},
\beq
\hat q(l)\equiv \frac{\partial \Delta k_T^2}{\partial l}.
\label{240}
\eeq
Calculated within the framework of the dipole approach \cite{zkl}, the broadening rate was found to be\cite{jkt},
\beq
\hat q(l)=2C\,\rho_A(l),
\label{260}
\eeq
where
\beq
C={1\over2}\,\vec\nabla_{r_T}^2 \sigma_{\bar qq}(r_T)\Bigr|_{r_T=0},
\label{280}
\eeq
and $\sigma_{\bar qq}(r_T)$ is the universal cross section dipole-proton cross section \cite{zkl}. Besides the transverse $\bar qq$ separation $r_T$ it also implicitly depends 
on the dipole energy \cite{broadening}. 
The dipole and hadronic cross sections are related as \cite{zkl},
\beq
\sigma_{hp}=\left\la\sigma_{\bar qq}(r_T)\right\ra \equiv
\int d^2r_T \left|\Psi_h(r_T)\right|^2
\sigma_{\bar qq}(r_T),
\label{290}
\eeq
where $\Psi_h(r_T)$ is the hadronic light-cone wave function.

Employing in (\ref{120}) the small-$r_T$ approximation, $\sigma_{\bar qq}(r_T)\approx Cr_T^2$, we can replace Eq.~(\ref{200}) by,
\beq
S^2(l_1,l_2)=\left\la
\exp\left[-{r_T^2\over2}\int\limits_{l_1}^{l_2}dl\, \hat q(l)\right]
\right\ra.
\label{300}
\eeq
Here instead of averaging the dipole cross section in the exponent, as was done in (\ref{200}),
we average the whole exponential. This way the Gribov corrections \cite{gribov,zkl} are included to all orders (except gluon shadowing, which comes from higher Fock components \cite{kst2}) in the eikonal Glauber expression (\ref{200}). 

Expression (\ref{300}) allows a direct generalization for absorption in a dense quark-gluon medium,
which is usually characterized with a transport coefficient $\hat q$. This will be done in the next section.

\subsection{Path integral description}\label{path}

In Eq.~(\ref{300}) the dipole transverse separation $r_T$ is assumed to be "frozen" by Lorentz time dilation, i.e. to remain unchanged during the time of propagation through the medium, which
might be the case only at sufficiently high dipole energies. The path integral technique presented below, allows to get rid of this approximation, and describe the propagation of a "breathing" dipole.

As was mentioned above, a $\bar cc$ pair produced in the c.m. frame of the colliding nuclei has a small initial separation $r_0\sim1/\sqrt{4m_c^2+p_T^2}$ and then propagates through
the medium with momentum $p_{\psi}\equiv p_T$. In such a kinematics the initial dipole size can never be "frozen" by Lorentz time dilation.
Indeed, the distribution amplitude of the produced $\bar qq$
contains different relative momenta of quarks. However, the large relative momentum components are heavy and get out of phase at long distances, where only small momentum, i.e. slowly expanding states are in coherence.
As a result, the mean dipole separation rises with path length as $\propto \sqrt{l}$, 
rather than linearly.
If a $\bar cc$ dipole is produced and moving with momentum $p_\psi$, its transverse (relative to its momentum) separation rises as \cite{kps-pT,knps},
\beq
r_T^2(l) = \frac{2l}{\alpha(1-\alpha)p_\psi}+r_0^2,
\label{140}
\eeq
where $\alpha$ is the fractional light-cone momentum of the quark. From this expression we conclude that the initial dipole size $r_0^2\sim1/p_\psi^2$ plays no role at a sufficiently long distance $l$, even if $p_\psi$ is large. Therefore, the initial dipole size cannot be "frozen".

While the relation (\ref{140}) is obviously oversimplified, a rigorous quantum-mechanical description of the propagation of a free colorless $\bar cc$ can be
performed with the path-integral method \cite{kz91,kst1}. One has to sum up all possible trajectories of propagation of quark and antiquark in order to
incorporate the effects of fluctuation of the dipole separation. 
In the light-cone variables the Green function of the dipole propagating in a medium satisfies
the 2-dimensional Schr\"odinger-type equation~\cite{kz91} (see detailed derivation in the Appendix of hep-ph/9808378 in \cite{kst1}) 
\beqn
&&i\frac{\partial}{\partial l}G\left(l,\vec r_T;0,\vec r_{0}\right) =
\nonumber\\ &&
\left[\frac{m_c^2-\Delta_{r_T}}{2E_\psi\alpha(1-\alpha)}
+U_{\bar cc}\left(r_T,l\right)\right]
G\left(l,\vec r_T;0,\vec r_{0}\right),
\label{320}
\eeqn
with the boundary condition,
\beq
G_{\bar cc}(l,\vec r_{T};0,\vec r_{0})\Bigr|_{l=0} =
\delta(\vec r_{T}-\vec r_{0}).
\label{340}
\eeq
Here $l$ is the longitudinal coordinate along the straight trajectory of the $\bar cc$ dipole
in the rest frame of the medium. The dipole evolves its transverse separation from the initial value $\vec r_0$ 
at $l=0$ up to $\vec r_T$ after propagating a distance $l$. To avoid possible confusions we remind that $r_T$ is transverse relative to the charmonium trajectory, which itself is orthogonal to the nuclear collision axis.

The real part of the effective potential  describes binding effects ${\rm Re}\,U(r_T)=V(r_T)$.
The Lorentz boost of this potential from the $\bar cc$ rest frame is a theoretical challenge, which we deal with below in section \ref{boost}. Here we avoid this problem, eliminating the real and keeping only the imaginary part of the potential, which is responsible for interaction of the dipole and the medium,
as was described above,
\beq
{\rm Im}\,U_{\bar cc}(r_T,l)=-{v_\psi\over4}\hat q(l)\,r_T^2,
\label{350}
\eeq 
where $v_\psi=p_\psi/E_\psi$ is the dipole velocity relative to the medium, $E_\psi=\sqrt{p_\psi^2+4m_c^2}$ is its energy.

The survival probability amplitude of a charmonium produced inside a hot medium is given by
the convolution of the Green function with the initial and final distribution amplitudes,
\beqn
\left|S\right|^2 = \left|\frac{\int d^2 r_1 d^2 r_2\Psi_{f}^\dagger(r_2)
G(L,\vec{r_2};0,\vec{r_1})
\Psi_{in}(r_1)}
{\int d^2 r\, \Psi_{f}^\dagger(r)\,\Psi_{in}(r)}\right|^2\!\!.
\label{400}
\eeqn
The final one $\Psi_{f}(r,\alpha)$ is the charmonium wave function, which is dominated by equal sharing of the light-come momentum, $\alpha\approx 1/2$ (see section \ref{boost}.

\section{Debye screening }\label{debye}

\subsection{Production time scales}

Production of a charmonium with momentum $p_\psi=p_T$  is characterized with two time scales \cite{brod-muel,kz91}.
Creation of a heavy quark pair takes a time interval, usually called coherence time,
\beq
t_c\sim\frac{2E_\psi}{4m_c^2+p_T^2}=\frac{2}{E_\psi},
\label{110}
\eeq
Notice that there is no naively expected Lorentz time dilation here, on the contrary, this time scale contracts with rising charmonium energy.

It takes much longer time for the quark pair to form the charmonium wave function. This time interval, usually called formation time, reads,
\beq
t_f\sim\frac{1}{\omega}\,\frac{E_\psi}{2m_c},
\label{120}
\eeq
where $\omega={1\over2}(m_{\psi'}-m_{J/\psi})$ is the oscillator frequency, which characterises the time $\sim1/\omega$ of circling over the orbit in the rest frame of a bound $\bar cc$ system. In the frame, where the charmonium is moving with energy $E_\psi$, the same time interval
 is Lorentz delayed, as given by (\ref{120}).
 
 \subsection{Maximal melting}\label{maximal}
 
  Intuitively is clear that if the binding potential is weakened (for whatever reason), or even completely  eliminated, for a short time interval $\Delta t\ll1/\omega$ in the charmonium rest frame, the bound state will not be strongly disturbed, because the quarks will have no time to move away from their orbits. In the medium rest frame, in which the $\bar cc$ pair is moving with velocity $v_\psi=p_\psi/E_\psi$, the same situation corresponds to a formation time, much longer than the propagation time in the medium,
\beq
t_f\gg R_A/v_\psi.
\label{130}
\eeq
We assume that the transverse dimensions of the medium are of the order of the nuclear radius. Thus, the Debye screening effect is expected to vanish for charmonia produced  in heavy ion collisions with large $p_T=p_{\psi}$.

\begin{figure}[htb]
\begin{center}
 \includegraphics[width=8cm]{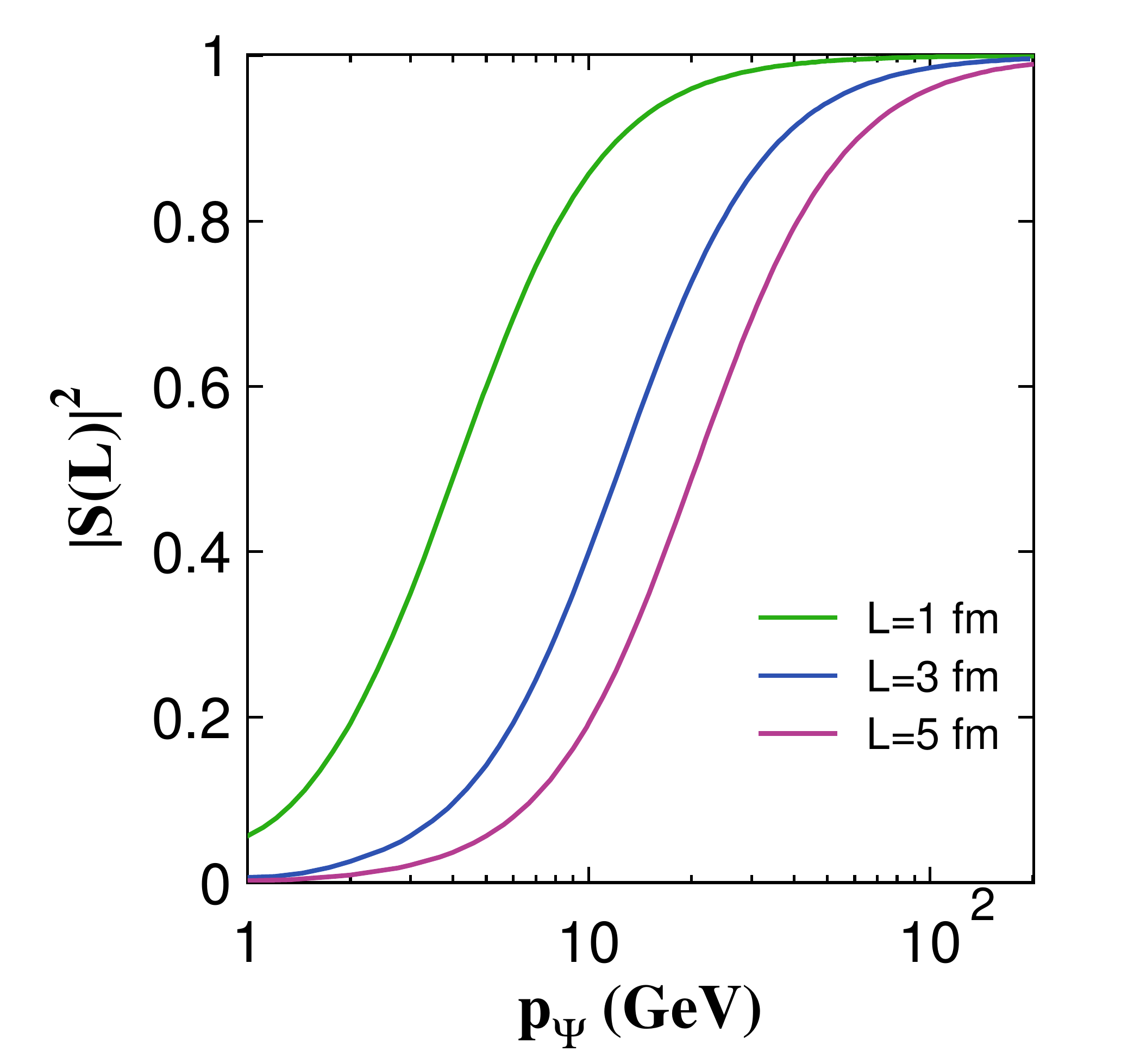}
 \caption{\label{fig:maximal} (Color online) The attenuation factor $\left|S(L)\right|^2$ for $J/\psi$ produced with momentum $p_{\psi}\equiv p_T$ off a hot medium. The Debye screening in the medium is assumed to have maximal magnitude completely eliminating the $\bar cc$ binding potential during in-medium propagation over the path length $L$,}
  \end{center}
 \end{figure}

Naively, one could think that the binding potential between $q$ and $\bar q$ is the reason
why a high-energy dipole does not expand quickly like in classical physics, $r_T\propto l$, and the potential prevents the quarks of flying away with the initial large relative momenta. This is not true: even lacking any interaction potential the $\bar qq$ dipole keeps a small separation 
over a long path length due to coherence. Indeed, the distribution amplitude of the produced $\bar qq$
contains different relative momenta of quarks. However, the large relative momentum components are heavy and get out of being in phase at long distances, where only small momenta, i.e. slowly expanding states are in coherence. 
As a result, the mean dipole separation rises with path length as $\propto \sqrt{l}$, 
rather than linearly, as was demonstrated in Eq.~(\ref{140}). These effects are fully included into the path-integral description Eq.~(\ref{400}), which, moreover, is capable to account for a binding potential. The latter possibility is investigated below, but at this point we want to demonstrate on an exaggerated example of  a binding potential, which completely vanishes
inside the medium (infinitely high temperature), that the survival probability of the dipole remains finite. Let us assume the following coordinate dependence of the potential (no time dependence),
\beqn
{\rm Re}\,U(r)=\left\{\begin{array}{cc}{\rm Re}\,U(r)=0\ \ \ \ \ \ \  & {\rm inside\ the\ medium};\\
{\rm Re}\,U(r)={\rm Re}\,U_{vac}(r) & {\rm outside\ the\ medium}.
\end{array}\right.
\nonumber
\eeqn
To clear up this example we simplify the scenario avoiding other sources of attenuation and inessential technical complications. In particular,  we eliminate the absorptive imaginary part of the potential, Eq.~(\ref{350}); keep the Debye screening maximally strong along the whole path length $L$ of the $\bar cc$ in the medium; and use the oscillator form of the real part of the potential outside the medium. Correspondingly both the initial and final wave functions in (\ref{200}) have the Gaussian form, $\Psi(r)=(\lambda/\sqrt{\pi})\,\exp(-\lambda^2r^2/2)$ with $\lambda^2=m_c^2+p_{\psi}^2/4$ and $\lambda^2=m_c\,\omega/2$ for initial and final size distribution functions respectively.
Then the equation (\ref{140}) can be solved analytically, and the suppression factor Eq.~(\ref{200}) for a $\bar cc$ propagating with velocity $v_\psi$ over the path length $L$ takes the form \cite{kz91},
\beq
\left|S(L)\right|^2=\frac{m_c^2\,p_\psi}{16\pi^2L}\,
\left(1+\frac{\omega}{2m_c}\right)
\,\left|I\right|^2,
\label{137}
\eeq
where
\begin{widetext}
\beq
\left|I\right|^2 =\Biggl|
\int\limits_{-\infty}^\infty dx_1dx_2
\exp\left\{-{1\over2}x_1^2\,m_c^2 
-{1\over4}x_2^2\,m_c\omega
-\frac{i\,p_{\psi}}{8L}(x_1-x_2)^2\right\}\Biggr|^2
=\frac{8\pi^2}{m_c^2}\left[\omega^2 m_c^2+
\frac{p_\psi^2}{4L^2}\left(1+\frac{\omega}{2m_c}\right)^2\right]^{-1/2}
\label{144}
\eeq
\end{widetext}
The resulting suppression factor $\left|S(L)\right|^2$ is plotted in Fig.~\ref{fig:maximal} as function of  $J/\psi$ momentum  for $L=1,\ 3,$ and $5\,\fm$.
Of course the $J/\psi$ survival probability vanishes towards $p_{\psi}=p_T=0$ because $c$ and $\bar c$ just fly away with no chance to meet, i.e. to be projected on the $J/\psi$ wave function. Later, however we will see that in reality even at small momenta $p_\psi$ the survival probability is fairly large  because of the rapid decrease of temperature in expanding plasma
and restoration of the binding potential.

In another limit of high momentum the survival probability approaches unity because the expansion of the initially small dipole size is slowed down by Lorentz time dilation. In this perturbative regime of full color transparency the presence of a binding potential plays no role.

Notice that in the real situation of heavy ion collisions the effective path length in a hot deconfined medium is rather short, 2-3 times shorter than the nuclear radius. This happens
because the medium density is falling with time,
and the dipoles produced in the central area have low chance to survive. Below we incorporate these features into the calculations. 

Meanwhile, this exaggerated example of a maximal screening (infinitely high temperature) demonstrates a sizeable
survival probability of $J/\psi$, even if no binding potential exists inside the medium.

\subsection{Lorentz boosted Schr\"odinger equation}\label{boost}

To proceed further we need a model for the biding potential, as well as for the screening corrections. For the former we employ the realistic Cornell form of the potential 
\cite{cornell,buchmueller}, which includes the Coulomb-like and the linear string behaviour of the potential at small and large separations respectively. However, such a 3-dimensional potential is appropriate only for a charmonium at rest, while the mean momentum squared of $J/\psi$ produced in heavy ion collisions at LHC varies from $7$ to $10\GeV^2$ \cite{psi-alice} in the collision c.m. frame. So one faces a challenge of Lorentz-boosting of the potential.

The same problem concerns the Debye screening effects, which are known well in the charmonium rest frame from the lattice simulations~\cite{Bazavov,Kaczmarek:2003ph,Kaczmarek:2012ne}. Again, these results cannot be applied to a $\bar cc$ pair moving with a relativistic velocity.

Although the Schr\"odinger equation on the light cone was derived and solved~\cite{BL,PirnerGalowSchlaudt,PirnerNurpeissov} the relation between potential in rest frame and light cone is a challenge. Recently, a nonlinear transformation of the potential, which allows to reproduce the mass spectrum, was proposed in~\cite{Trawinski:2014msa}.
However, as is demonstrated below, one can
get a correct spectrum with a linear transformation of the potential. 
We also derive an equation of internal motion of heavy quarks system propagating with a relativistic velocity.

In the large-$m_{c}$ limit the intrinsic motion of 
a system of heavy quarks is non-relativistic~\cite{Korner:1991kf}
and the Fock space of charmonium is dominated by the lowest $\bar{c}c$ state,
whose dynamics can be described in terms of an effective potential.
In what follows, we use a standard heavy quark mass expansion. In this limit, the binding energy is small, so the difference of momenta carried by the quark and antiquark is also small,
\cite{psi-kth},
\beq
\la\lambda^2\ra\equiv\left\la \left(\alpha-1/2\right)^2\right\ra=
\frac{\la p_L^2 \ra}{4m_c^2}=
{1\over4}\la v_L^2\ra,
\label{220}
\eeq
where $\alpha$ is the quark fractional light-cone momentum; $p_L$ and $v_L$ are the longitudinal momentum and velocity of the charm quark in the rest frame of the charmonium. The mean quark (3-dimensional) velocity, evaluated with realistic potential \cite{psi-kth,hikt1} is $\la v^2\ra\approx 0.2$, so the mean value in (\ref{220})
is rather small, $\la\lambda^2\ra=0.017$.

\begin{figure}[htb]
\begin{center}
 \includegraphics[width=5cm]{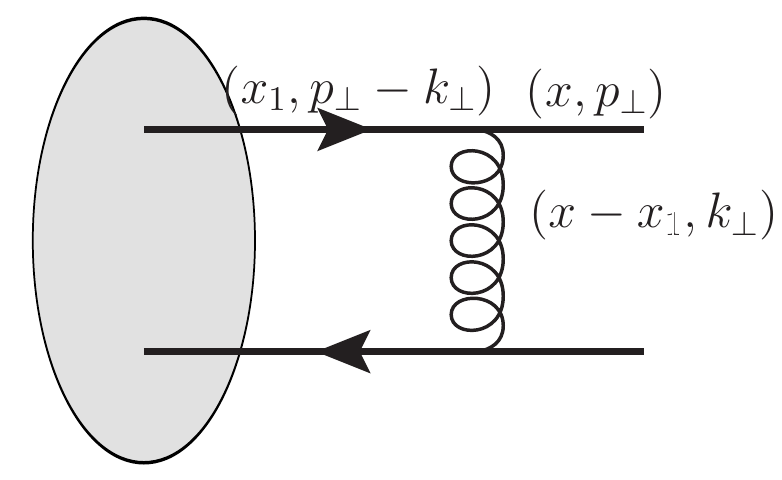}
 \caption{\label{fig:BS} Kinematic variables for the Bethe-Salpeter bound state diagram.}
  \end{center}
 \end{figure}

Here we present a brief description of the transformation procedure based on smallness of $\la\lambda^2\ra$. A detailed derivation will be published elsewhere~\cite{KLSS}.
Let us assume that in the $\bar cc$ rest frame $V(r)$ is the interaction 
potential and its Fourier image $\tilde{V}\left(\vec{k}\right)$
corresponds to the propagator of a hypothetical vector particle exchange in $t$-channel.
The Bethe-Salpeter equation for the eigenvalues is shown
schematically in the Fig.~\ref{fig:BS}
Making expansion over
the small parameter $\lambda$, Eq.~(\ref{220}), we arrive at the equation of motion~\cite{KLSS},
\begin{eqnarray}
 &  & p_{\psi}^{-}\phi\left(\alpha,r_{\perp}\right)=\\
 &  & \frac{-\Delta_{\perp}+m_c^{2}}{2p_{\psi}^{+}\, \alpha(1-\alpha)}\phi\left(\alpha,r_{\perp}\right)+\hat{V}\phi\left(\alpha,\, r_{\perp}\right),\nonumber 
\end{eqnarray}
where $p_{\psi}^{\pm}$ are the components of the charmonium light-cone momentum.The interaction term here reads,
\begin{equation}
\hat{V}\phi\left(\alpha,r_{\perp}\right)=\int\limits_0^1 d\beta\, 
K\left(\alpha,\beta,r_{\perp}\right)\phi\left(\beta,r_{\perp}\right),
\label{2020}
\end{equation}
with the kernel $K\left(\alpha,\beta,r_{\perp}\right)$ given by 

\begin{eqnarray}
 &  & K\left(\alpha,\beta,r_{\perp}\right)\approx\label{eq:K}\\
 &  & \approx\frac{1}{\pi}\frac{M_{\psi}}{p_{\psi}^{+}}
 \int\limits_{r_{\perp}}^{\infty}dr\, r\, V(r)\frac{\cos\left(2m_{c}\left|\alpha-\beta\right|\sqrt{r^{2}-r_{\perp}^{2}}\right)}{\sqrt{r^{2}-r_{\perp}^{2}}}\nonumber \\
 &  & =\frac{1}{2\pi}\frac{M_{\psi}}{p_{\psi}^{+}}
 \int\limits_{0}^{\infty}dz\, V\left(\sqrt{r_{\perp}^{2}+z^{2}}\right)
 e^{2im_{c}(\alpha-\beta)z},\nonumber 
\end{eqnarray}
where $M_{\psi}$ is the charmonium mass. Since the kernel~(\ref{eq:K})
depends only on the difference between the fractional light-cone momenta $\alpha-\beta$, it is convenient to work with a Fourier transform $\tilde{\phi}$ defined as
\begin{equation}
\tilde{\phi}\left(\zeta,\, r_{\perp}\right)=
\int\limits_{0}^{1}\frac{d\alpha}{2\pi}\,\phi\left(\alpha, r_{\perp}\right)\, 
e^{2im\left(\alpha-\frac{1}{2}\right)\zeta}.
\label{2040}
\end{equation}
In this case the variable $\zeta$ has a meaning of the minus-component
of the light-cone separation between quarks; the factor $2m_{c}$ is introduced
to keep the correct dimension and will help to simplify some of the formulas
below. Expanding (\ref{eq:K}) over a small parameter $\lambda$ and replacing $\lambda\Rightarrow i\partial \zeta/2m$ in coordinate space, we get the final result for the Schr\"odinger equation \cite{KLSS},
\begin{eqnarray}
 \frac{M_{\psi}^{2}}{2p_{\psi}^{+}}\,\tilde{\phi}\left(\zeta,\, r_{\perp}\right)&=&
 \frac{-\Delta_{\perp}-\partial_{\zeta}^{2}+m_c^{2}}{p_{\psi}^{+}/2}\tilde{\phi}\left(\zeta, r_{\perp}\right)
\nonumber  \\& + &
 \frac{M_{\psi}}{p_{\psi}^{+}}\,
 V\left(\sqrt{r_{\perp}^{2}+\zeta^{2}}\right)\tilde{\phi}\left(\zeta,r_{\perp}\right),
 \label{final}
\end{eqnarray}
which is explicitly invariant under Lorentz boosts. If we substitute
now $M_{\psi}\approx2m_c+\epsilon_{\psi}$, where $\epsilon_{\psi}$
is the binding energy of the charmonium, in the charmonium rest frame in 
the leading order in $\epsilon_{\psi}$,
we obtain the usual nonrelivistic Schr\"odinger equation.

Now we are in a position to replace Eq.~(\ref{320}), in which the longitudinal motion of quarks was frozen, by an equation with the kinetic and potential terms, which describes
the 3-dimentional motion of the quarks,
  \beqn
\Biggl[i\frac{\partial}{\partial z_+}\!\!&+&\!\!2\,
\frac{\Delta_{r_T}+(\partial/\partial\zeta)^2 -m_c^2}{p_\psi^+}
-U_{\bar cc}\left(\zeta,\vec r_T\right)\Biggr]\nonumber\\ &\times&
G\left(z^+,\zeta,\vec r_T;z_1^+,\zeta_1,\vec r_{1T}\right)=0.
\label{322}
\eeqn
Here the real part of the boosted potential ${\rm Re}\,U(\zeta,\vec r_T)$ is related to the rest frame binding potential $V(r)$ as,
\beq
{\rm Re}\,U(\zeta,\vec r_T)=V\left(r=\sqrt{r_T^2+\zeta^2}\right),
\label{324}
\eeq
while the imaginary part of the potential remains the same as in Eq.~(\ref{350}),
${\rm Im}\,U_{\bar cc}(r_T,l)=-\hat q(l)\,r_T^2/4$. 

The real and imaginary parts of the potential are related to different parameters characterizing the medium, the temperature and transport coefficient, respectively. To proceed further we need to establish a relation between them via the equations of state. At large $T\gg T_{c}$, where $T_{c}$ is the critical temperature, this relation is simple,
$\hat{q}\approx3.6\, T^{3}$, but at lower temperatures $T\lesssim T_{c}$,
it is is more complicated and is parametrized in \cite{Chen:2010te}.

\section{Numerical results}\label{results}

\subsection{Attenuation of $J/\psi$} \label{j/psi}

Now we are able to solve numerically Eq.~(\ref{322}) for the Green function describing the propagation of a colorless $\bar cc$ pair through a hot medium, employing realistic potentials and screening corrections, although they are known only in the charmonium rest frame.
As a model for the biding potential, we employ the realistic Cornell form of the potential \cite{cornell}, which successfully reproduces the spectrum of charmonia.
For medium screening effects we rely on the popular approach, which is 
based on the Debye-H\"uckel theory and the lattice results, parameterized in the analytic form proposed in\emph{~}\cite{Karsch:1987pv},

\beq
V_{\bar cc}\left(r,\, T\right) =\frac{\sigma}{\mu(T)}
\left(1-e^{-\mu(T)r}\right)-\frac{\alpha}{r}e^{-\mu(T)r},
\label{1040}
\eeq
where
\beqn
\mu(T) & =&g(T)T\,\sqrt{\frac{N_{c}}{3}+\frac{N_{f}}{6}},
\nonumber\\
g^{2}(T)&=&\frac{24\pi^{2}}{33\ln\left(19T/\Lambda_{\bar{MS}}\right)};
\eeqn
$\sigma=1\GeV/\!\fm$ is the string tension;
and the parameter $\alpha\approx0.471$. The bound states in such a potential
are terminated at high temperatures, when the Debye radius $r_{D}=1/\mu(T)$ becomes smaller than the mean radius of the charmonium. We use this bound state potential in the real part of the boosted potential, Eq.~(\ref{324}), of equation (\ref{322}).

The imaginary part of the boosted potential, Eq.~(\ref{350}) is controlled by the transport coefficient $\hat q$, varying with coordinates and time, for which
we employ the popular model \cite{Chen:2010te},
\beq 
\hat q(t,\vec b,\vec\tau)=\frac{
q_0\,t_0}{t}\, \frac{n_{part}(\vec b,\vec\tau)}{n_{part}(0,0)}
\,\Theta(t-t_0),
\label{1060} 
\eeq
Here $\vec b$ and $\vec\tau$ are the impact parameter of the collision and the point, where the medium is observed, respectively. The medium density is assumed to be proportional to the number of participants  $n_{part}(\vec b,\vec\tau)$, defined in the standard way \cite{klaus}. The 
falling time dependence $1/t$ is due to longitudinal expansion.  It is also assumed 
that the medium takes time $t_0$ to reach equilibrium, and in (\ref{1060}) is assumed that the medium "appears" at this moment. Of course such a treatment of the medium development is grossly oversimplified, and the time scale $t_0$ is poorly defined. In what follows we fix $t_0=1\fm$.

The maximal value of $\hat q$ at $\vec b=\vec s=0$ and $t=t_0$ is treated as 
a  free parameter, to be adjusted to data, which can vary depending on the particular nucleus and  collision energy. This is the usual strategy of such an analysis: to probe the medium properties with production of charmonia.

After Eq.~(\ref{322}) is solved and the Green function is found, one can calculate the 
survival probability factor $S^2(b)$ of a charmonium produced in a collision of nuclei $A$ and $B$ with relative impact parameter $b$ (compare with (\ref{400})),
\begin{widetext}
\beq
\left|S(b)\right|^2=
\int\limits_0^{2\pi}\frac{d\phi}{2\pi}
\int \frac{d^2s\,T_A(\vec s)T_B(\vec b-\vec s)}{T_{AB}(b)}
\left|\frac{\int d^2r_1 d^2r_2 d\zeta_1 d\zeta_2\Psi_f^\dagger(\vec r_2,\zeta_2)
G(\infty,\vec r_2,\zeta_2;0,\vec r_1,\zeta_1)
\Psi_{in}(\vec r_1,\zeta_1)}
{\int d^2r d\zeta\,\Psi_f^\dagger(\vec r,\zeta)\,\Psi_{in}(\vec r,\zeta)}\right|^2.
\label{330}
\eeq
\end{widetext}
Here we assume that the $\bar cc$ dipole propagates along a straight trajectory, which starts at impact parameter $\vec s$ relative to the nucleus $A$ center. $\phi$ is the azimuthal angle between this trajectory and the scattering plane (vector $\vec b$). 
The initial distribution function in (\ref{330}) is assumed to have a Gaussian form,
\beq
\Psi_{in}(\vec r_T,\zeta)=\frac{a}{\sqrt{\pi}}e^{-a^2r^2/2},
\label{1020}
\eeq
with $r=\sqrt{r_T^2+\zeta^2}$, and $a\sim1/\sqrt{m_c^2+p_T^2}$.
Since the initial expansion of a small dipole is very fast,  the original separation is not important.

The Green function $G(\infty,\vec r_2,\zeta_2;0,\vec r_1,\zeta_1)$, describing the dipole evolution, and the final wave function $\Psi_f(\vec r_2,\zeta_2)$,
are the solutions of Eqs.~(\ref{final}) and (\ref{322}) respectively, with the same boosted potential (\ref{324}). This is important for self-consistency of the method and for the correct  behavior of Eq.~(\ref{322}) in the limit of a dilute medium.

\begin{figure}[htb]
\begin{center}
 \includegraphics[width=7cm]{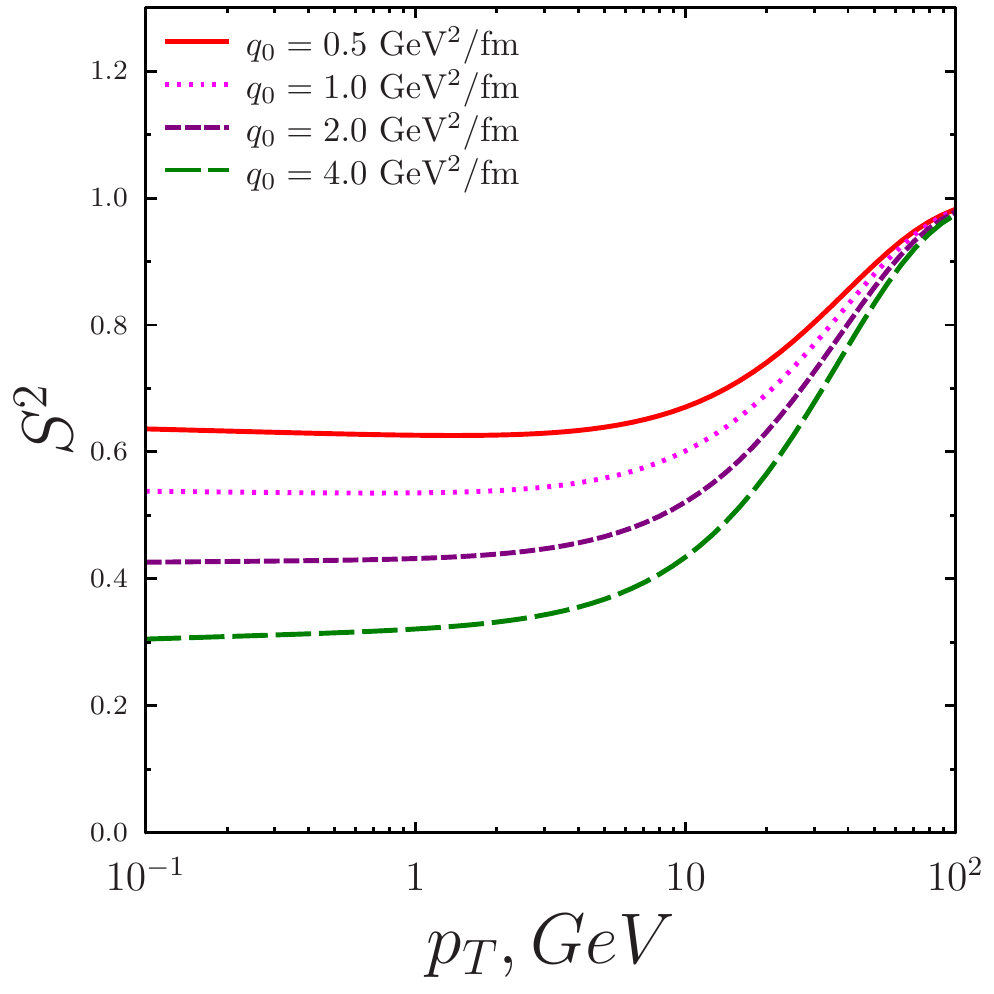}
 \caption{\label{fig:melting} (Color online) The FSI suppression factor calculated with Eq.~(\ref{330}) for $J/\psi$ produced with momentum $p_{\psi}\equiv p_T$ in central lead-lead collisions. The absorption effects are eliminated by fixing ${\rm Im}\,U=0$, in order to single out the net effect of Debye screening.
 The curves from top to bottom are calculated with $q_0=0.5,\ 1,\ 2$ and $4\GeV^2\!/\!\fm$.}
  \end{center}
 \end{figure}

In order to disentangle the effects of melting and absorption, we would like to look at each of them in a pure  form, i.e. eliminating the other one.
First of all, to see the net melting effect we solve Eq.~(\ref{322}) excluding absorption, i.e. fixing ${\rm Im}\, U=0$. The results for $\left|S(b=0)\right|^2$, calculated with Eq.~(\ref{330}) for a central lead-lead collision,  are plotted in Fig.~\ref{fig:melting}  for different trial values of the parameter $q_0=0.5,\ 1,\ 2$ and $4\GeV^2\!/\!\fm$.
Notice that similar to Fig.~\ref{fig:maximal} demonstrating the maximal melting effect, 
the survival probability $|S|^2$ rises with $p_\psi$ and saturates at unity. This happens for the same reason: Lorentz time dilation slows down the expansion of the initially small size of the $\bar cc$ dipole, and color transparency~\cite{zkl} makes the medium more transparent. In such a perturbative regime screening of the binding potential, even existence of any potential, plays no role. The charmonium wave function is formed outside, far away from the medium. 

On the contrary, the behavior at small $p_\psi\to0$ in Fig.~\ref{fig:melting} is different from the case of maximal melting effect, Fig.~\ref{fig:maximal}. 
Although a deconfined $\bar cc$ pair tends to fly away with no chance to meet again,
in the more realistic case considered here,  the transport coefficient Eq.~(\ref{1060})
is finite and is falling with time as $1/t$. Therefore the medium cools down and the binding potential is restored, so that the colorless dipole has a non-zero chance to survive.
 
\begin{figure}[htb]
\begin{center}
 \includegraphics[width=7cm]{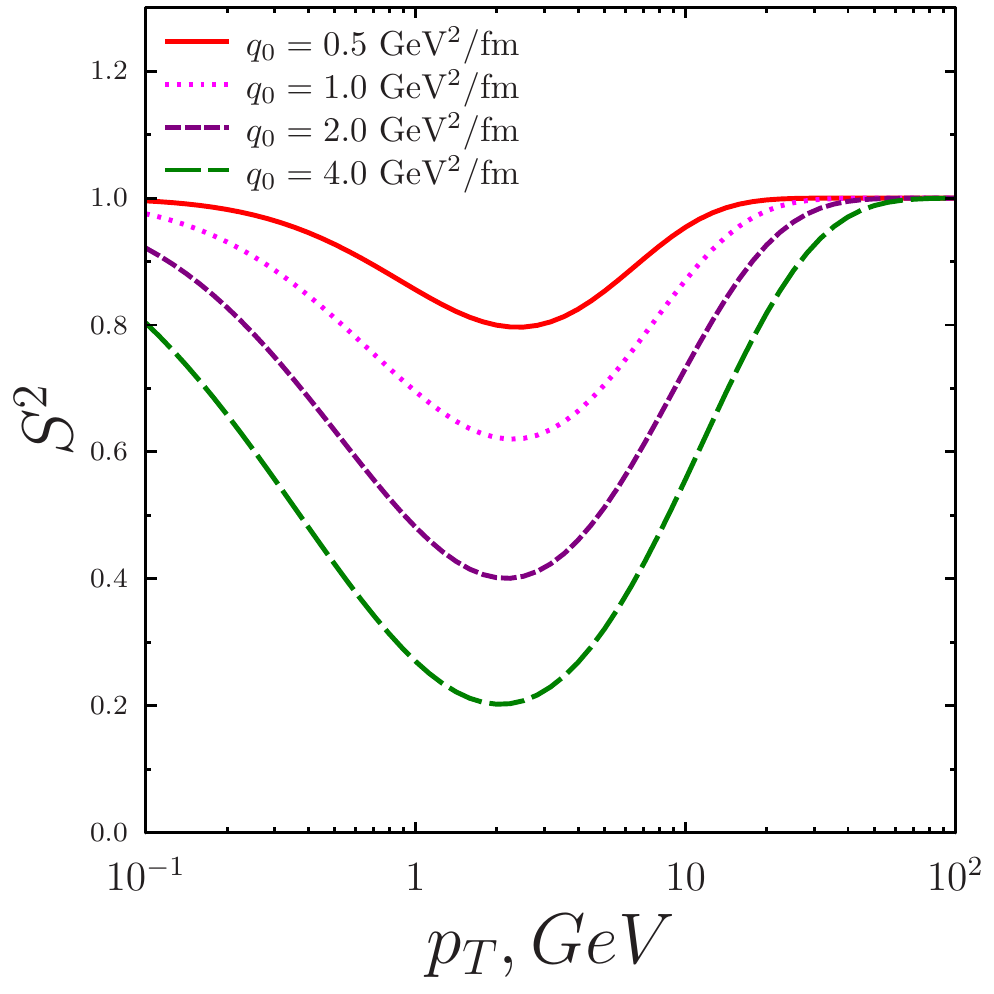}
 \caption{\label{fig:absorption} (Color online) Same as in Fig.~\ref{fig:melting}, but with the net effect of absorption, i.e. with eliminated Debye screening.}
  \end{center}
 \end{figure}

The other extreme  is to eliminate the melting effect by fixing in (\ref{1040}) $T=0$, but include absorption introducing the imaginary part of the potential according to Eq.~(\ref{350}). In this way one can see a net effect of absorption. The results, using Eq.~(\ref{330}), are plotted in Fig.~\ref{fig:absorption} for different values of $q_0$.
Again, the survival probability approaches unity at high $p_\psi=p_T$. for the same reason as in
the previous examples, due to color transparency. A peculiar feature of the observed dependence on $p_\psi$ is in this case
a similar behavior as before, but in the opposite limit of $p_\psi\to0$., which is caused by a combined effect
of two factors. First, the absorption rate Eq.~(\ref{350}) is proportional to the velocity of the dipole, so it vanishes in this limit. However, a dipole even with a small velocity will eventually propagate through the whole path length in the medium, and the accumulated
amount of the passed medium is velocity independent. This would be the case in a static medium, however the absorption rate Eq.~(\ref{350}) is  proportional to $\hat q$, which
is falling with time according to Eq.~(\ref{1060}). Thus, eventually absorption vanishes at $p_\psi\to0$.

\begin{figure}[htb]
\begin{center}
 \includegraphics[width=7cm]{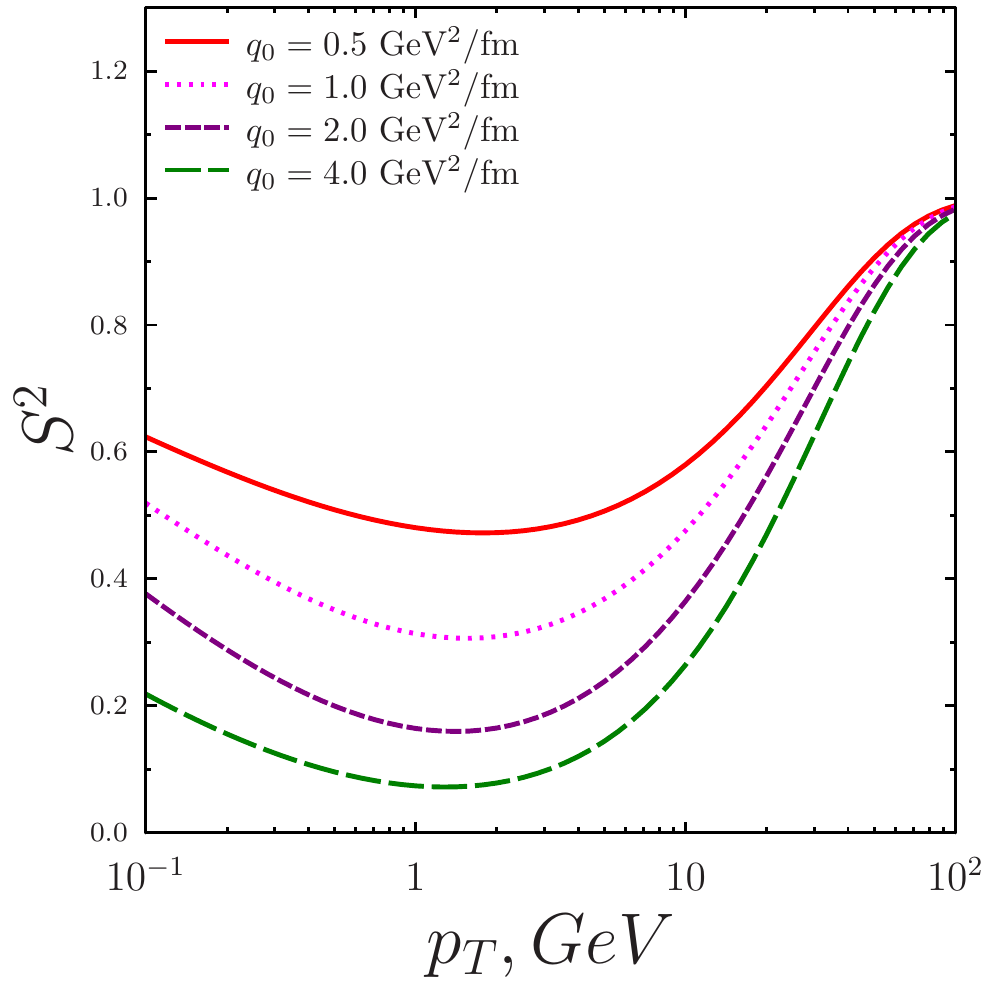}
 \caption{\label{fig:full} (Color online) Same as in Figs.~\ref{fig:melting} and \ref{fig:absorption}, but including both, screening and absorption effects.}
  \end{center}
 \end{figure}

The full results incorporating both effects of melting and absorption cannot be simply deduced from the above two extreme cases of net melting or net absorption effects, because they strongly correlate.
For example, absorption leads to a color filtering effect, a stronger attenuation of large dipoles.
As a result, the mean dipole size decreases and the Debye screening becomes weaker.
Solving Eq.~(\ref{322}) with a full potential containing the real, Eq.~(\ref{324}), and imaginary, Eq.~(\ref{350}), parts, and averaging over impact parameters with Eq.~(\ref{330})
we arrive at the final results for the suppression factor $|S|^2$ plotted in Fig.~\ref{fig:full}.

\begin{figure}[htb]
\begin{center}
 \includegraphics[width=7cm]{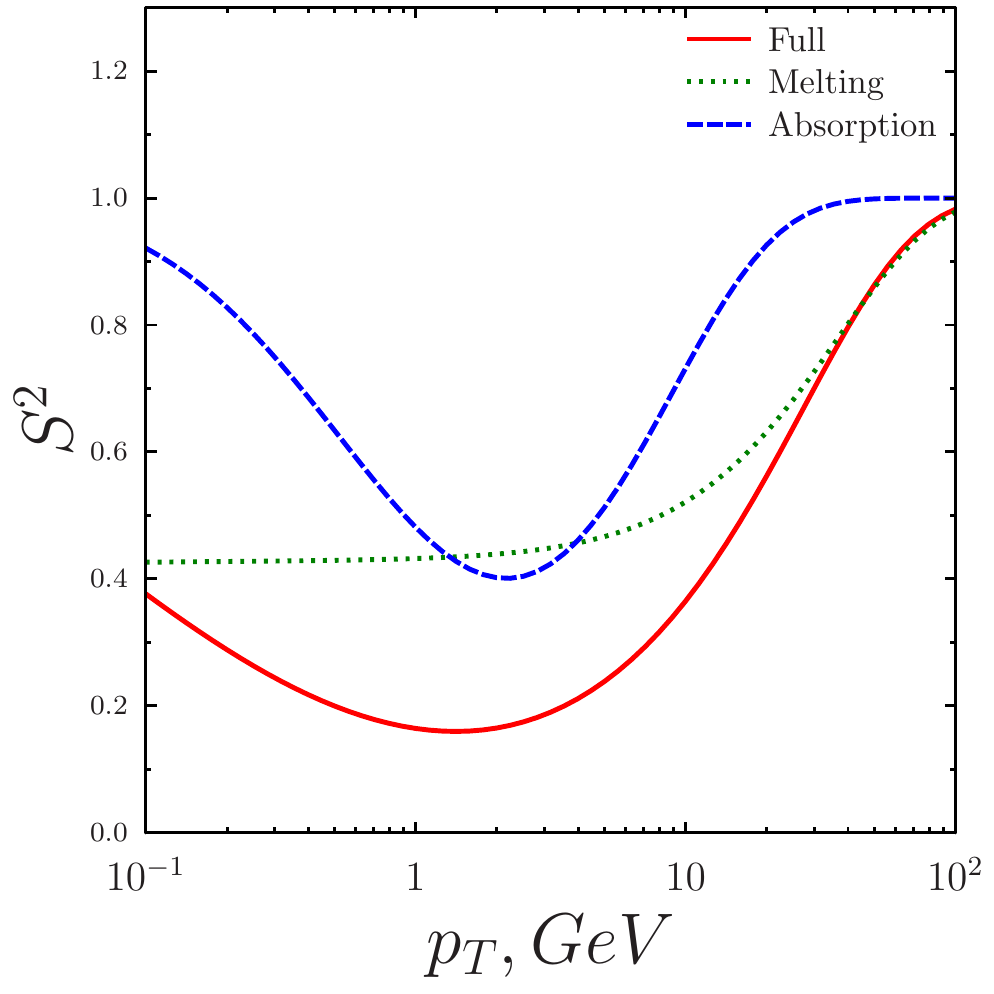}
 \caption{\label{fig:3curves} (Color online) The curves from top to bottom show the attenuation factor $\left|S(L)\right|^2$ vs $J/\psi$ momentum for either pure absorption, or pure melting, and for both effects included, respectively. The calculations are performed for central lead-lead collisions with $q_0=2\GeV^2\!/\!\fm$. }
  \end{center}
 \end{figure}

Eventually, for comparison, we plot all three versions presented in Figs.~\ref{fig:melting}-\ref{fig:full} on one plot Fig.~\ref{fig:3curves} for the selected value of $q_0=2\GeV\!/\!\fm$,
which looks to us realistic, since was extracted from the analysis \cite{knps}
of data for high-$p_T$ hadron suppression in lead-lead collisions at LHC.
We see that the magnitude of suppression caused by the two mechanisms is similar at transverse $J/\psi$ momenta around the mean value.
The total suppression effect is maximal at $p \sim\la p_\psi\ra$ and is rather strong,
about $|S|^2\approx0.2$.

\subsection{Attenuation of $\psi(2S)$}\label{psi'}

The first radial excitation $\psi(2S)$ is known to have the mean radius squared about twice as big as $J/\psi$. Correspondingly, a stronger melting effect has been always expected for 
$\psi(2S)$. Apparently, the absorption effects should be stronger than for $J/\psi$ as well.
The suppression factor $|S|^2$ is given by the same Eq.~(\ref{330}), but the final wave function $\Psi_f$ is a solution of Eq.~(\ref{2040}) for the radial excitation $\psi(2S)$.
\begin{figure}[htb]
\begin{center}
 \includegraphics[width=7cm]{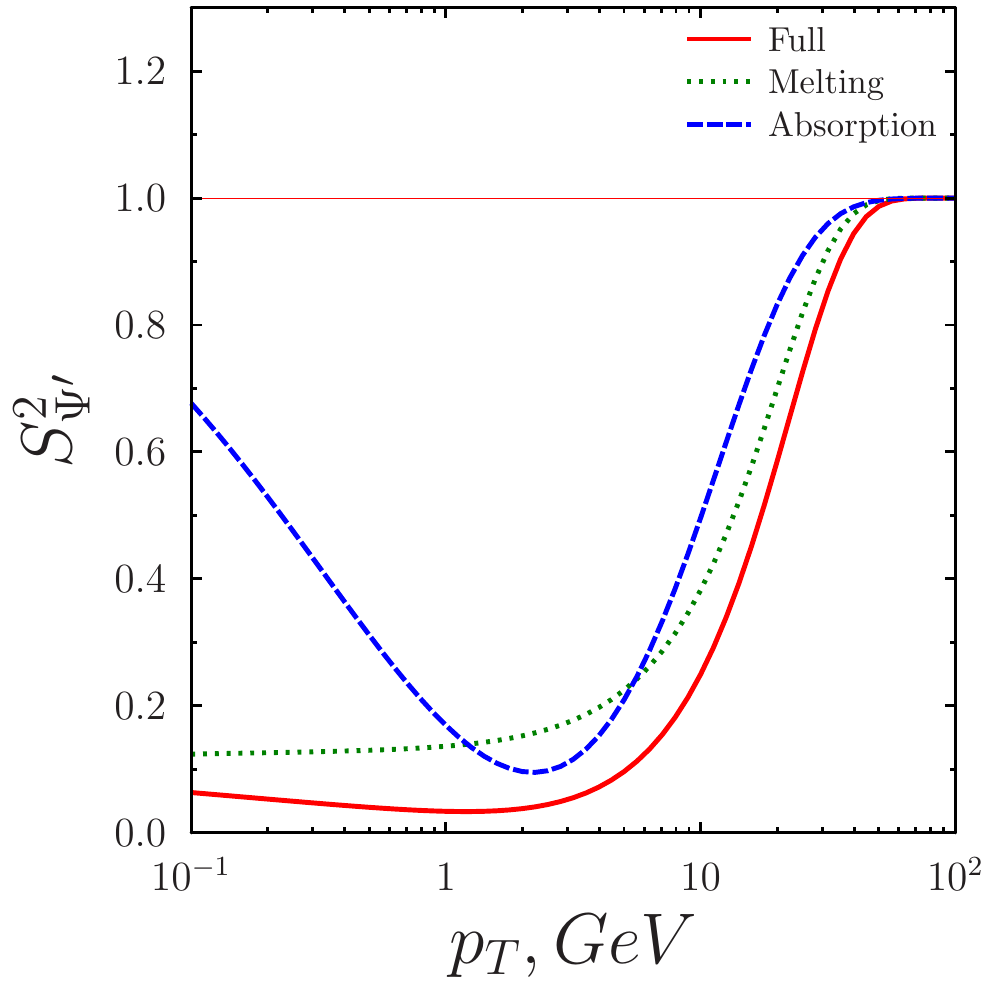}
 \caption{\label{fig:psi(2s)} (Color online) The same as in Fig.~\ref{fig:3curves}, but for production of a $\psi(2S)$ radial excitation.}
  \end{center}
 \end{figure}
The results of calculations with $q_0=2\GeV^2\!/\!\fm$ are shown in Fig.~\ref{fig:psi(2s)}
for the suppression factors for net melting (dotted), net absorption (dashed) and full effect (solid).
As expected,  suppression  depicted in Fig.~\ref{fig:psi(2s)} is considerably stronger  compared with $J/\psi$, Fig.~\ref{fig:3curves}. This is confirmed by the $\psi(2S)$ to $J/\psi$ ratio plotted in 
Fig.~\ref{fig:2s-1s}, which is indeed mostly below unity.
\begin{figure}[htb]
\begin{center}
 \includegraphics[width=7cm]{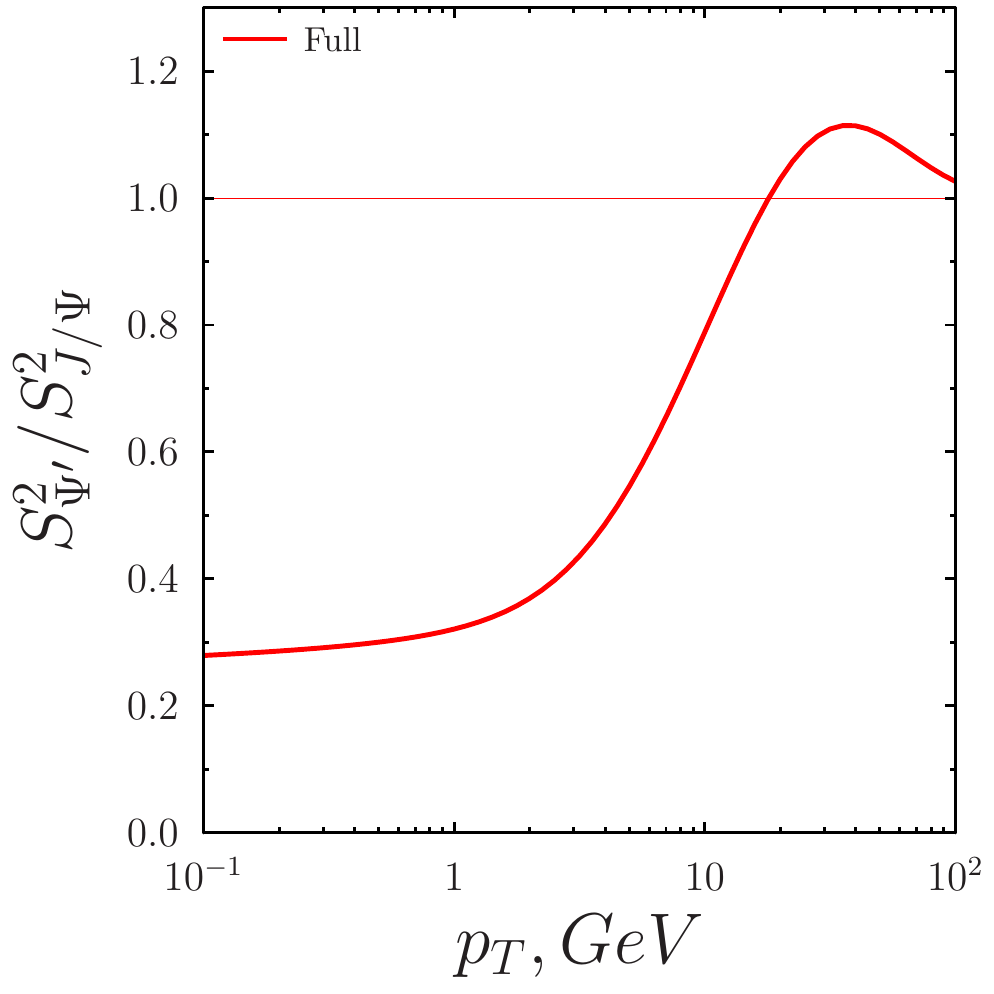}
 \caption{\label{fig:2s-1s} (Color online) Ratio of production rates of $\psi(2S)$ to $J/\psi$ calculated for central lead-lead collisions with $q_0=2\GeV^2\!/\!\fm$.  }
  \end{center}
 \end{figure}

However, this ratio also demonstrates a peculiar feature: at large $p_\psi\gsim10\GeV$.
It exceeds unity, at variance with the naively expected stronger in-medium attenuation of $\psi(2S)$. This is related to the effect of nuclear enhancement of the $\psi(2S)$ photo-production rate, predicted long time ago in  \cite{kz91,knnz}. This effect is explained by the specific structure of the $\psi(2S)$ wave function $\Psi_{\psi}(r)$, which has a node and changes sign at $r\approx0.4\fm$ (see Fig.~4 in \cite{ihkt1}). This leads to a considerable compensation in the overlap between the initial $\bar cc$ distribution function and the wave function of $\psi(2S)$.
Due to color filtering, leading to a shrinkage of the initial distribution, the projection to the $\psi(2S)$
wave function increases, which results in an enhancement. This interesting effect
is expected to occur at sufficiently high $p_\psi$.

\subsection{Azimuthal asymmetry}

\begin{figure}[htb]
\begin{center}
 \includegraphics[width=7cm]{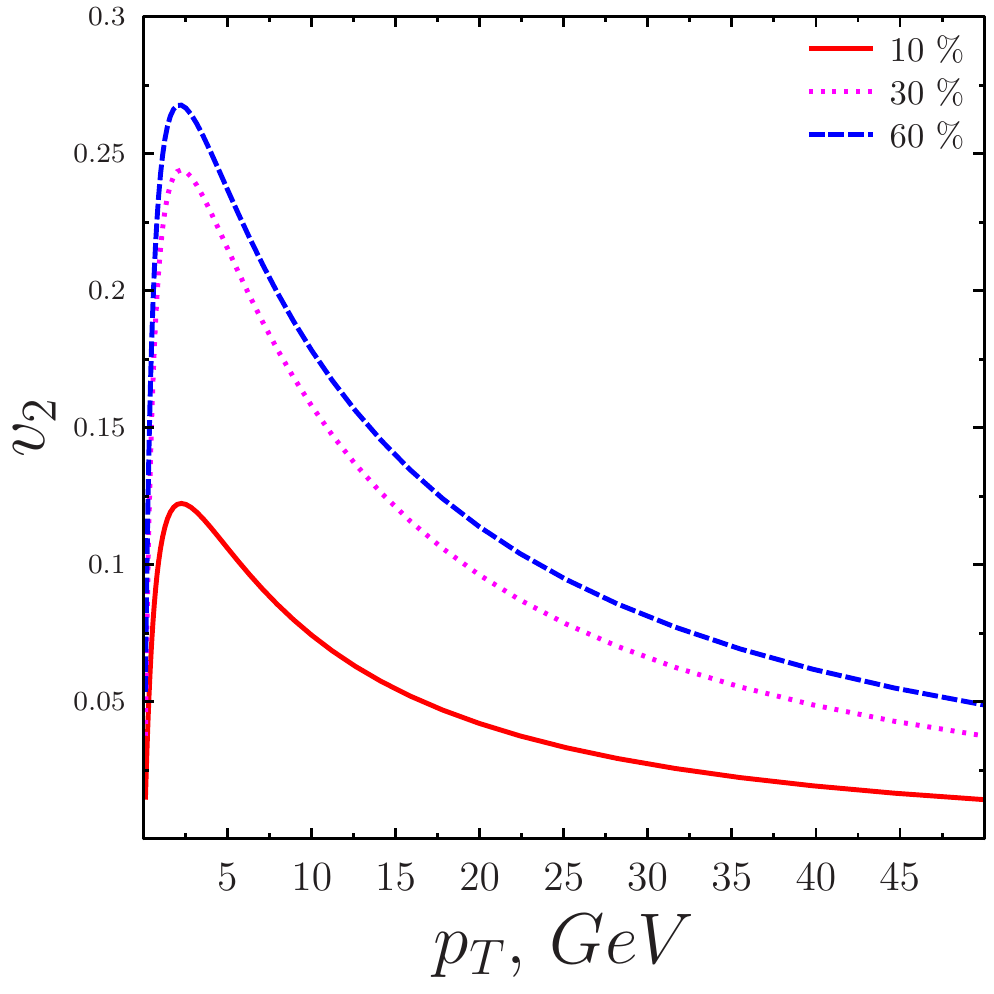}
 \caption{\label{fig:v2} (Color online) Azimuthal asymmetry parameter $v_2$ for $J/\psi$ production in lead-lead collisions calculated with $q_0=2\GeV^2\!/\!\fm$.
  The curves represent different centralities, from bottom to top $10\%,\ 30\%$ and $60\%$. }
  \end{center}
 \end{figure}

A strong charmonium attenuation in the final state medium should lead to an azimuthal asymmetry, characterised by the parameter $v_2=\la\cos(2\phi)\ra$, where $\phi$ defined in (\ref{330}), is the azimuthal angle between the $J/\psi$ trajectory and the scattering plane. Indeed, the overlap between two nuclei colliding with impact parameter $b>0$, and the transverse spacial distribution of the produced dense matter, 
have elliptic shapes. Apparently, a charmonium produced along the long or short diameters of the nuclear intersection area, propagates different path lengths in the medium, so is suppressed differently, which gives rise to an azimuthal asymmetry. Correspondingly, we can calculate $v_2(b)$ as,
\begin{widetext}
\beq
v_2(b)=
\int\limits_0^{2\pi}\frac{d\phi}{2\pi}\,\cos(2\phi)
\int d^2s\,\frac{d^2s\,T_A(\vec s)T_B(\vec b-\vec s)}{T_{AB}(b)|S_{J/\psi}(b)|^2}
\left|\frac{\int d^2r_1 d^2r_2 d\zeta_1 d\zeta_2\Psi_f^\dagger(\vec r_2,\zeta_2)
G(\infty,\vec r_2,\zeta_2;0,\vec r_1,\zeta_1)
\Psi_{in}(\vec r_1,\zeta_1)}
{\int d^2r d\zeta\,\Psi_f^\dagger(\vec r,\zeta)\,\Psi_{in}(\vec r,\zeta)}\right|^2.
\label{1200}
\eeq
\end{widetext}
The results obtained with $q_0=2\GeV^2\!/\!\fm$, are plotted in Fig.~\ref{fig:v2} as function of $J/\psi$ momentum for several centralities. 
Naturally, $v_2$ vanishes at very large momenta, where the medium becomes fully transparent, as was demonstrated if Figs.~\ref{fig:melting}-\ref{fig:3curves}.
The parameter $v_2$ at intermediate and large momenta has a magnitude similar to what was measured at LHC and calculated in \cite{knps} for light hadrons. It is also close to what was observed for $J/\psi$ by the CMS experiment recently \cite{cms-v2-psi}. It is worth emphasizing
that the azimuthal asymmetry is not affected by ISI, because the expected effects \cite{amir} are 
very weak.
So, on the contrary to the attenuation factor $S^2(b)$, the calculated $v_2$ could be directly compared with data, provided that we did not miss any significant FSI effect.
The latter does not seem to be the case. Indeed, recent data from LHC give a strong indication about the presence  of a mechanism enhancing $J/\psi$, at least at low $p_T$ (see more in the next section).
Therefore, we restrain here from direct comparison with data.

\section{Summary}\label{summary}

The main objectives of this paper are: (i) to improve the available theoretical tools for evaluation of the effect of Debye screening of the charmonium binding potential in a hot medium; (ii) to draw attention to the importance of the usually missed alternative mechanism of charmonium attenuation, which is absorption, i.e. interaction of the $\bar cc$ dipole with the medium.
Correspondingly, we performed the following studies.

\begin{itemize}

\item
Fig.~\ref{fig:maximal} demonstrates that even in the limit of infinitely high temperature,
when Debye screening completely eliminates the binding potential, the survival probability of a charmonium is not zero. Moreover, it approaches unity at large transverse momenta of the charmonium.

\item
The effect of the temperature dependent Debye screening has been calculated so far on the lattice only for a charmonium embedded at rest in a hot medium. We developed a procedure of Lorentz boosting of the binding potential and screening corrections, into a reference frame moving with a finite velocity. The boosted Schr\"odinger equation has the form of Eqs.~(\ref{final})-(\ref{322}). In these equations the real part of the boosted potential is directly related to the rest frame binding potential by Eq.~(\ref{324}), where the longitudinal quark separation is replaced by the variable $\zeta$ defined in (\ref{2040}).

\item
Another source of charmonium attenuation in the medium is the possibility of color-exchange interactions
of the $\bar cc$ pair with the medium, which break-up the colorless dipole.
Such interaction, which we call absorption, contributes to the imaginary part of the potential
according to Eq.~(\ref{350}). The effect of absorption is subject to color transparency, therefore it vanishes at large charmonium momenta.

\item
Solving the equation (\ref{322}) with a realistic model for the coordinate and time dependences of the temperature and transport coefficient, we calculated the suppression factor
for a $J/\psi$ produced in central lead-lead collisions for the cases of a net melting effect
(Fig.~\ref{fig:melting}), net absorption (Fig.~\ref{fig:absorption}), and including both effects  (Fig.~\ref{fig:full}). We found that the two mechanisms of meting and absorption 
give comparable contributions to the $J/\psi$ suppression. Similar calculations and conclusions (but stronger suppression) were done for the radial excitation $\psi(2S)$ (Fig.~\ref{fig:psi(2s)}).

\item
Attenuation of a charmonium produced in nuclear collisions unavoidably leads to an azimuthal
asymmetry for non-central collisions. We calculated the parameter $v_2$ (Fig.~\ref{fig:v2})
and found it to have a magnitude comparable with data.
\end{itemize}

\section{Outlook}\label{outlook}

We do not compare the results of calculations with data, because some important mechanisms
involved in charmonium production off nuclear collisions are still missing.

\begin{itemize}

\item
Initial state interactions (ISI), sometimes labelled as cold nuclear matter effects, essentially affect the charmonium production rate. Its magnitude cannot be deduced from data on
$pA$ collisions for many reasons \cite{nontrivial}. 

First, the ISI and FSI stages do not factorise.
Indeed, in the c.m. of nuclear collision the ISI stage lasts a very short time $\Delta t\sim
R_A 2m_N/\sqrt{s}\sim 10^{-3}\fm$. The time needed to form the charmonium wave function is much longer, according to Eq.~(\ref{120}) $t_f\sim 1/\omega\approx0.7\fm$ (at $p_\psi=0$). So the ISI stage ends up with production
of a small-size $\bar cc$ pair, rather than a charmonium, differently from $pA$ collisions.
This is why we used such a $\bar cc$ dipole with the distribution function Eq.~(\ref{1020}),
rather than the $J/\psi$ wave function.

Second, the ISI stage in $AA$ collisions includes novel effects compared to $pA$ mechanisms of $\bar cc$ dipole interactions, such as the effects of double color filtering and of mutual boosting
of the saturation scales \cite{nontrivial,rhic-lhc,boosting,puzzles}
 
 \item
 The FSI absorption is controlled by the transport coefficient $\hat q$. Contrary to the  temperature, the transport coefficient, defined in  (\ref{240}) as the rate of broadening,
is not only a characteristic of the medium, but also of the  interaction. It is known from data and from theory~\cite{broadening} that $\hat q$ of a cold nuclear matter is a steeply rising function of  energy. Therefore, the effect of FSI absorption should become stronger at larger $p_\psi$.
This fact should be taken into consideration upon comparison with data, which we avoid in the present study. One may wonder how the temperature, which is essentially a feature of the medium, can be related to $\hat q$, which is controlled by the interaction. The answer is that
such a relation, established in \cite{bdmps}, is relevant only to the broadening calculated
in the Born approximation. The higher order corrections, corresponding to multi-gluon radiation, which generate the rising energy dependence of broadening \cite{broadening},
break down the relation between $T$ and $\hat q$. These corrections are to be added to the 
starting value of the $\hat q$ related to the temperature. 

\item
The exponential attenuation of a colorless dipole related to the absorption rate Eq.~(\ref{350})
exaggerates the magnitude of suppression. There is a strict bottom bound for the survival probability of a colorless dipole propagating through a dense medium, $|S|^2 \geq 1/N_c^2=1/9$. Indeed, as a result of  many color-exchange interactions in a medium a $\bar cc$ pair becomes completely unpolarized in color space, i.e. all $N_c^2$ possible color states of this pair must be equally presented. The colorless state is only one of these $N_c^2$, so its probability is as small as $1/N_c^2=1/9$ and cannot be less than that \cite{alesha}.
Therefore our results, which show a comparable or stronger than $1/N_c^2$ suppression in a  dense medium, should be corrected for the effect of regeneration
of a colorless state due to color octet-to-singlet transitions. Notice that such an effect of  regeneration explains \cite{hkz} the observed unusual nuclear enhancement of $J/\psi$ photoproduction.

\item
The ISI stage can end up not only with the production of a colorless $\bar cc$ pair, but a color-octet $\bar cc$ can also be produced with a much higher probability. While in $pA$ collisions such a channel does not contribute to the charmonium production, in $AA$ collisions
the FSI stage, which starts with a color-octet state, can end up with a color singlet $\bar cc$ dipole. This happens due to color-exchange interactions with the medium, which lead to regenerating octet-to-singlet transitions \cite{alesha,psi-kth}. Thus, our present description of the FSI stage
which assumes a colorless dipole as an input, is not complete and should be supplemented with such a novel mechanism of regeneration of colorless dipole from the original color octet state
\cite{novel}.  

\item
The rate of broadening is related to the transport coefficient of the medium only if the former is calculated in the lowest order Born approximation \cite{bdmps}. However, broadening
rises with parton energy due to higher order corrections, which is well confirmed by data \cite{broadening}. So far every calculation of jet quenching and other hard probes, including charmonium suppression presented in this paper, have been done assuming that the transport coefficient $\hat q$ is independent of $p_T$. We plan to improve the present results implementing a $p_T$-dependent $\hat q$.

\end{itemize}


\begin{acknowledgments}\vspace{-5mm}
We are grateful to Genya Levin for many useful discussions, and to Miguel Escobedo
for providing missed references.
This work was supported in part
by Fondecyt (Chile) grants 1130543, 1130549, 1140390, 
and 1140377.
\end{acknowledgments}

\vspace*{2cm}

\end{document}